\documentclass[notitlepage, final, 10pt, pra, aps, twocolumn, floatfix, letter, longbibliography, english, superscriptaddress]{revtex4-2}

\pdfoutput=1

\usepackage[T1]{fontenc}
\usepackage[utf8]{inputenc}
\usepackage[english]{babel}
\usepackage{lmodern}

\usepackage{amsfonts}
\usepackage{amssymb}
\usepackage{amsmath}
\usepackage{bm}
\usepackage{bbm}
\usepackage{cases}
\usepackage{graphicx}
\usepackage[english]{babel}
\usepackage{xcolor}
\definecolor{navyblue}{rgb}{0.0, 0.0, 0.5}
\usepackage{latexsym}
\usepackage{booktabs}
\usepackage[protrusion=true, expansion=true]{microtype}
\usepackage[colorlinks=true, breaklinks=false, linkcolor=navyblue, urlcolor=navyblue, citecolor=navyblue]{hyperref}
\usepackage{float}

\usepackage{mathtools}
\usepackage{braket}
\usepackage{siunitx}
\newcommand{\ketE}[3]{\ket{E_{#1 #2}^{(#3)}}}
\newcommand{\braE}[3]{\bra{E_{#1 #2}^{(#3)}}}
\newcommand{\pr}{\hat{\Pi}}
\newcommand{\nop}{\hat{n}}
\newcommand{\ham}{\hat{H}}
\newcommand{\lyxmathsym}[1]{\ifmmode\begingroup\def\b@ld{bold}
\text{\ifx\math@version\b@ld\bfseries\fi#1}\endgroup\else#1\fi}

\DeclareMathOperator{\Tr}{Tr}

\makeatother

\begin{document}

\author{Oksana Busel}
\affiliation{Nano and Molecular Systems Research Unit, University of Oulu, Oulu 90014, Finland}
\author{Sami Laine}
\affiliation{Nano and Molecular Systems Research Unit, University of Oulu, Oulu 90014, Finland}
\affiliation{Department of Information Technology, Oulu University of Applied Sciences, Oulu 90101, Finland}
\author{Olli Mansikkam\"{a}ki}
\affiliation{Nano and Molecular Systems Research Unit, University of Oulu, Oulu 90014, Finland}
\author{Matti Silveri}
\affiliation{Nano and Molecular Systems Research Unit, University of Oulu, Oulu 90014, Finland}
\date{\today}

\title{Dissipation and Dephasing of Interacting Photons in Transmon Arrays}

\begin{abstract}
Transmon arrays are one of the most promising platforms for quantum information science. Despite being often considered simply as qubits, transmons are inherently quantum mechanical multilevel systems. Being experimentally controllable with high fidelity, the higher excited states beyond the qubit subspace provide an important resource for hardware-efficient many-body quantum simulations, quantum error correction, and quantum information protocols. Alas, dissipation and dephasing phenomena generated by couplings to various uncontrollable environments yield a practical limiting factor to their utilization. To quantify this in detail, we present here the primary consequences of single-transmon dissipation and dephasing to the many-body dynamics of transmon arrays. We use analytical methods from perturbation theory and quantum trajectory approach together with numerical simulations, and deliberately consider the full Hilbert space including the higher excited states. The three main non-unitary processes are many-body decoherence, many-body dissipation, and heating/cooling transitions between different anharmonicity manifolds. Of these, the many-body decoherence -- being proportional to the squared distance between the many-body Fock states -- gives the strictest limit for observing effective unitary dynamics. Considering experimentally relevant parameters, including also the inevitable site-to-site disorder, our results show that the state-of-the-art transmon arrays should be ready for the task of demonstrating coherent many-body dynamics using the higher excited states. However, the wider utilization of transmons for ternary-and-beyond quantum computing calls for improving their coherence properties.
\end{abstract}

\maketitle

\section{Introduction}
Transmon arrays have recently taken substantial advances in size, coherence, and controllability, opening doors for exciting demonstrations of quantum information protocols~\cite{Ofek16, Rosenblum18, Hu19, Arute2019, CI20, Wu21, Chen21, Marques21, Gong21, Krinner22, Zhao22, Acharya22, Chen22} and many-body simulations~\cite{Roushan2017, Ma2019, HF20, Carusotto2020, Guo2021, Mi21, Satzinger21, Blok21, Zanner22, Braumuller2022, Morvan22, Mi22, Zhu22, Saxberg22}. Transmons are typically operated as quantum two-level systems, qubits, despite their inherent nature as anharmonic oscillators with approximately $d\sim 5-10$ well-defined quantum states~\cite{Koch07}. Treating them as proper quantum multilevel systems can be leveraged in several ways, including enhanced hardware efficiency and functionality of quantum error correction~\cite{Muralidharan17, Elder20}, fault-tolerant protocols~\cite{Campbell14, Rosenblum18}, and versatile quantum simulations with less mapping overhead~\cite{Orell2019, Wang20, Macdonell21, Olli2021, Olli2022}. As a result, the utilization of the higher excited states has recently garnered substantial interest, witnessed through the demonstrations of qutrit operations and algorithms with transmons~\cite{Peterer15, Zhang19, Morvan21, Blok21, Steinmetz22, Cervera-lierta22, Cao22, Roy22, Luo22, Goss22}, other superconducting quantum devices~\cite{Neeley09, Kononenko21}, as well as trapped ion and photonic platforms~\cite{chi22, Ringbauer22}.

Taking the higher excited states into account, a transmon array can be accurately described using the Bose-Hubbard model with attractive interactions~\cite{deGrandi15, Orell2019, Roushan2017, Carusotto2020}. In our previous works, we have derived an effective model for the unitary dynamics of highly-excited states of coupled transmons based on nearly-degenerate perturbation theory~\cite{Olli2021,Olli2022}. In the typical parameter regime where the transmon anharmonicity $U$ dominates the hopping rate $J$, these states can be interpreted as quasiparticles exhibiting, for example, edge localization and effective long-range interactions. This has since found an application in explaining emergent soliton dynamics~\cite{Soliton22}. In addition to the unitary dynamics of systems of transmons, dissipation and dephasing rates of the higher excited states of individual transmons have also been quite well characterized~\cite{Peterer15, Morvan21, Blok21}. However, the combination of these two topics \---- a quantitative understanding of the non-unitary dynamics for the higher excited states in transmon arrays \---- has not been studied before in detail.

In this work, we include the dissipation and dephasing processes into the many-body dynamics of transmon arrays both analytically and numerically. We identify three main processes, listed here in descending order of their typical effective rates: many-body decoherence, many-body dissipation, and heating/cooling induced by the combination of pure transmon dephasing processes and many-body dynamics.

The Bose-Hubbard model with attractive interactions conserves the total boson number, meaning that the unitary dynamics neither adds nor removes excitations from the system. This is broken by the many-body dissipation process inducing transitions between the different boson-number manifolds and occurring at a rate proportional to the instantaneous total boson number. The many-body decoherence process, on the other hand, reduces the coherence of many-body superpositions at a rate proportional to the squared distance between the many-body Fock states. Finally, in the parameter regime of strongly interacting bosons, $U/J\gg 1$, the many-body spectrum of the Bose-Hubbard model is split into well-separated regions with almost-conserved interaction energy~\cite{Olli2021, Olli2022}. The transmon dephasing process combined with the many-body dynamics breaks this quasi-conserved symmetry by inducing heating and cooling transitions between these so-called anharmonicity manifolds.

Our results show that, with experimentally realistic values for dissipation and dephasing, it should be possible to observe the many-body dynamics of the higher excited states between coupled transmons in state-of-the-art transmon arrays, even including the inevitable site-to-site disorder. We clearly see that the dephasing of the highly-excited states is one of the critical factors in their wider utilization in ternary-and-beyond quantum computation and simulations.

The article is organized as follows. In Sec.~\ref{sec:intro2}, we introduce the attractive Bose-Hubbard model of a transmon array, together with the typical dissipation and dephasing processes in transmons through a master equation formalism. Sections~\ref{sec:diss} and~\ref{sec:dephasing} focus individually on the effects of dissipation and dephasing processes. In Section~\ref{sec:disorder}, we describe phenomena induced by disorder in the dissipation and dephasing rates between the individual transmons. Conclusions and future outlook are presented in Sec.~\ref{sec:conc}.

\section{Open many-body dynamics in~a~transmon~array}
~\label{sec:intro2}
\begin{figure}[t]
\includegraphics[width=\columnwidth]{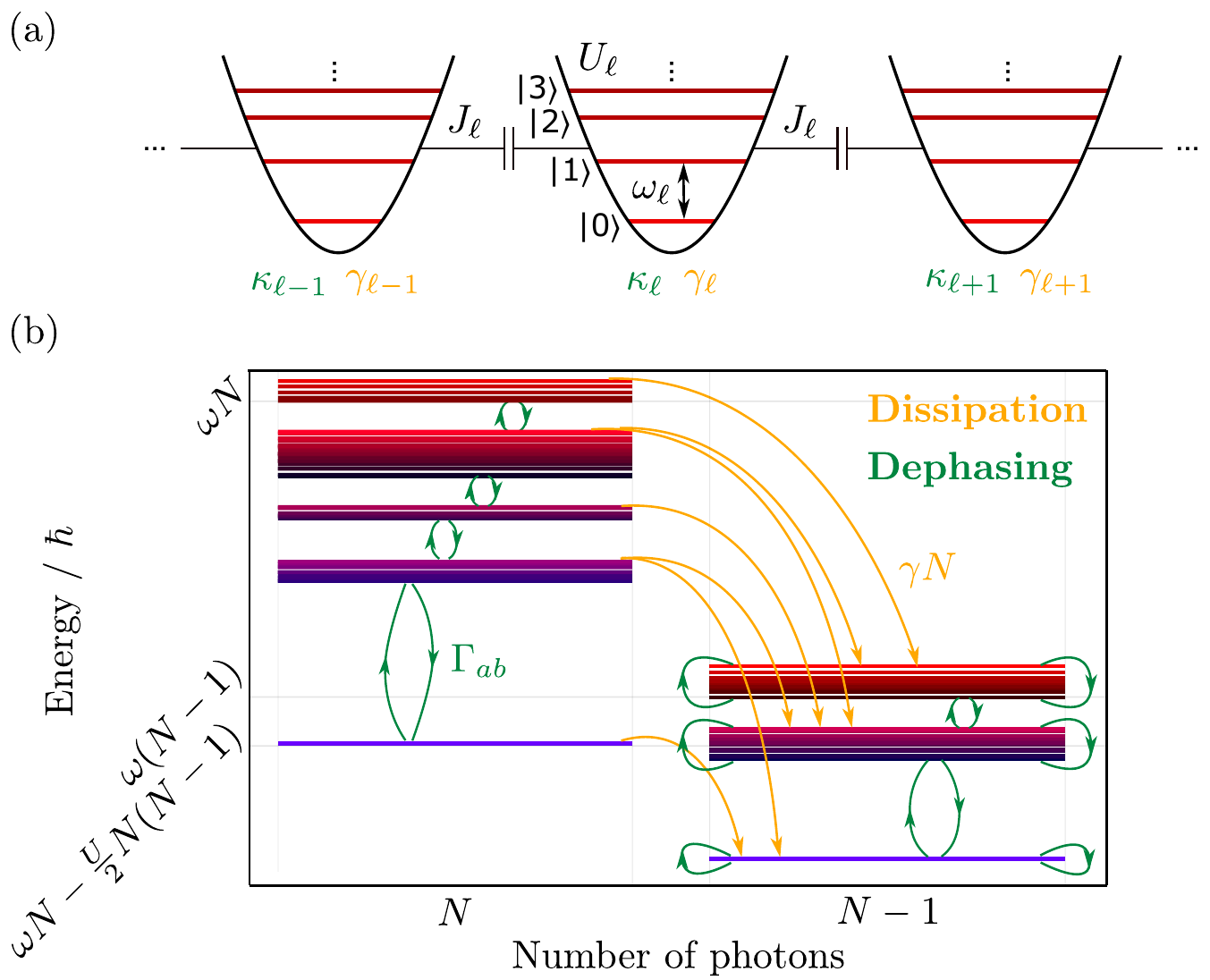}
\caption{(a) A schematic of a 1D transmon array, where the transmons are represented as anharmonic oscillators with frequencies~$\omega_\ell$, anharmonicities~$U_\ell$, nearest-neighbor hopping rates~$J_\ell$, dissipation rates~$\gamma_\ell$, and dephasing rates~$\kappa_\ell$. (b)~A~many-body energy level spectrum of a transmon array and a schematic showing many-body transitions due to the dissipation~(yellow) and dephasing~(green) processes. The colored bands denote the anharmonicity manifolds containing several many-body eigenstates. The red and blue colors of the energy levels represent the relative contributions of the hopping energy and anharmonicity, respectively.}
\label{fig:dndbhm-border-gradients-12-01-2022}
\end{figure}

A transmon is made of Josephson junctions and capacitor plates, realizing an anharmonic oscillator with natural frequency $\omega$ and anharmonicity $U$, see Fig.~\ref{fig:dndbhm-border-gradients-12-01-2022}(a). Nearby transmons interact with each other through a capacitive interaction $J$. In many-body language, the anharmonicity~$U$ describes the strength of the on-site many-body interactions between bosonic excitations, while $J$ is the hopping rate between neighboring transmons. Hence, an array of $L$ transmons is effectively described by the Bose-Hubbard model with attractive interactions~\cite{Orell2019,Carusotto2020},
\begin{align}
    \frac{\hat{H}_{\rm BH}}{\hbar}  =  \sum_{\ell=1}^{L}\omega_{\ell} \hat{n}_{\ell}  & -  \sum_{\ell=1}^L\frac{U_\ell}{2} \hat{n}_{\ell} \left( \hat{n}_{\ell}-1 \right) \notag \\ & + \sum_{\ell=1}^{L-1} J^{}_\ell \left(\hat{a}_{\ell}^{\dagger} \hat{a}^{}_{\ell+1}+\hat{a}^{}_{\ell} \hat{a}^{\dagger}_{\ell+1}\right),
    \label{eq:FullHam}
\end{align}
written here in the basis of the local bosonic annihilation $\hat{a}_{\ell}$, creation $\hat{a}_{\ell}^{\dagger}$, and occupation number $\hat{n}_{\ell} = \hat{a}_{\ell}^{\dagger} \hat{a}_{\ell}$ operators, with the reduced Planck’s constant $\hbar$.  The bosonic excitations described by the model are microwave photons but we use the term 'boson' for generality in this article. In modern arrays of transmons~\cite{Ma2019, Zhu22, Morvan22}, we have $\omega_\ell / 2 \pi \sim \SI{5}{\giga\hertz}$ for the on-site energies, $J_\ell / 2 \pi \sim \SIrange{10}{30}{\mega\hertz}$ for the hopping frequencies, and $U_\ell / 2 \pi \sim \SIrange{200}{250}{\mega\hertz}$ for the on-site interactions.Due to the inevitable small differences in manufactured devices, the parameters of any two transmons are usually not equal. However, for the sake of simplicity, we will assume in most parts of this work that there is no disorder in the parameters of the Hamiltonian, i.e., $\omega_\ell = \omega$, $J_{\ell} = J$, and $U_\ell = U$.

The transmon anharmonicity dominates the hopping frequency, $U \gg J$, resulting in an energy spectrum where states with the same total anharmonicity~$\hat A=-\sum_\ell \hat{n}_\ell (\hat{n}_\ell - 1) / 2$ form well-separated bands, see Fig.~\ref{fig:dndbhm-border-gradients-12-01-2022}(b). Due to the conservation of energy, the unitary dynamics of the model takes place mostly within the anharmonicity manifold of the initial state. The ratio~$J/U \ll 1$ can be considered a perturbation parameter. A highly-excited transmon can then be seen as a bosonic excitation -- a quasiparticle -- located at one site of the transmon chain and interacting with other quasiparticles, single bosons, or array edges~\cite{Olli2022}. The term `hard-core bosons' refers to the state manifold where the value of the total anharmonicity equals zero, having no excitations beyond the qubit subspace and having the highest energy, see Fig.~\ref{fig:dndbhm-border-gradients-12-01-2022}(b).

Transmons experience non-unitary dissipation and dephasing processes. The master equation yielding the non-unitary evolution of the density matrix of the system is given by
\begin{align}
    \frac{d \hat{\rho}}{d t} = - \frac{i}{\hbar}[ \hat{H}_{\rm BH}, \hat{\rho}] &+ \sum_{\ell=1}^{L} \frac{{\gamma}_{\ell}}{2} ( 2 \hat{a}_{\ell} \hat{\rho} \hat{a}_{\ell}^{\dagger} - \hat{a}_{\ell}^{\dagger} \hat{a}_{\ell} \hat{\rho} - \hat{\rho} \hat{a}_{\ell}^{\dagger} \hat{a}_{\ell})\nonumber\\
      &+  \sum_{\ell=1}^{L} \frac{\kappa_{\ell}}{2} ( 2 \hat{n}_{\ell} \hat{\rho} \hat{n}_{\ell} - \hat{n}^2_{\ell} \hat{\rho} - \hat{\rho} \hat{n}^2_{\ell}).
      \label{eq:MEF}
\end{align}
In transmon arrays, typical experimental values~\cite{Ma2019, Arute2020,  Gong21, Zhao22, Zhu22,  Saxberg22} for the mean dissipation rates $\gamma$ and the mean dephasing rates $\kappa$ are $\gamma / 2 \pi \sim \SIrange{5}{10}{\kilo\hertz}$ ($T_1~\sim \SIrange{15}{30}{\micro\second}$) and $\kappa / 2 \pi \sim \SIrange{50}{300}{\kilo\hertz}$ ($T^\star_2\sim\SIrange{1}{6}{\micro\second}$). In quantum information setups, where the devices are better isolated, one can achieve much better values~\cite{Place2021, Krinner22} $T_1~\sim \SIrange{30}{300}{\micro\second}$ and $T_2~\sim\SIrange{50}{100}{\micro\second}$. Here, we use the more conservative values from the array setups. The rates usually differ quite significantly from site to site, with typical standard deviations of $\sigma_\gamma/\gamma \sim \SIrange{0.2}{0.5}{}$ and $\sigma_\kappa/\kappa \sim \SIrange{0.3}{0.5}{}$. 
Despite the disorder, we first assume that all the rates are identical, $\gamma_\ell = \gamma $ and $\kappa_\ell = \kappa$, and then in Sec.~\ref{sec:disorder} shortly discuss the effects of the disorder. The higher excited states of a transmon have pronounced susceptibility to charge noise~\cite{Koch07}. This means that in practice, their dephasing rates are larger than implied by Eq.~\eqref{eq:MEF}, see, e.g., Refs.~\cite{Peterer15, Morvan21, Blok21}. We first consider the simple dephasing model of Eq.~\eqref{eq:MEF} and discuss its extension in Sec.~\ref{sec:disorder}. In what follows, we consider the many-body dynamics under dissipation and dephasing separately.

\section{Dissipation} \label{sec:diss}
Let us first focus on uniform dissipation ($\gamma_\ell=\gamma$) considering the master equation \begin{align}
    \frac{d \hat{\rho}}{d t} &= - \frac{i}{\hbar}[ \hat{H}_{\rm BH}, \hat{\rho}] +\sum_{\ell=1}^{L} \frac{\gamma}{2} \left( 2 \hat{a}^{}_{\ell} \hat{\rho} \hat{a}_{\ell}^{\dagger} - \hat{a}_{\ell}^{\dagger} \hat{a}^{}_{\ell} \hat{\rho} - \hat{\rho} \hat{a}_{\ell}^{\dagger} \hat{a}^{}_{\ell} \right)\notag\\
    & = -\frac{i}{\hbar}\left(\hat{H}^{}_{\rm NJ}\hat\rho-\hat\rho\hat{H}_{\rm NJ}^\dag\right)+\sum_{\ell=1}^L\gamma\hat{a}^{}_{\ell} \hat{\rho}  \hat{a}_{\ell}^{\dagger}.
   \label{eq:MEDiss}
\end{align}
On the second line, we have split the equation into two parts according to the quantum trajectory approach~\cite{Daley14}. There, the non-unitary dynamics is described by the no-jump evolution under the non-Hermitian Hamiltonian
\begin{equation}
    \hat H^{(\gamma)}_{\rm NJ}=\hat H_{\rm BH}-i \sum_{\ell=1}^L\frac{\hbar\gamma}{2}\hat a_{\ell}^\dagger\hat a_{\ell}^{}, \label{eq:diss_NJ}
\end{equation}
interrupted by random quantum jumps via the jump operators~$\sqrt{\gamma}\hat a_\ell$. The rate at which a particular quantum jump occurs is $\braket{\psi(t)|\gamma\hat a_\ell^\dagger a_\ell^{}|\psi(t)}=\gamma\braket{\hat n_\ell}$. The total rate is then given by $\braket{\psi(t)|\sum_\ell \gamma\hat a_\ell^\dagger a_\ell^{}|\psi(t)}=\gamma N$. Here, the instantaneous jump events remove a photon at the site $\ell$ and change the state discontinuously,
\begin{equation}
    \ket{\psi_{\rm QJ}(t)}= \frac{ \sqrt{\gamma} \hat{a}_{\ell} \ket{\psi(t)}} {\sqrt{ \braket{ \psi(t) | \gamma \hat{a}_\ell^{\dagger} \hat a^{}_\ell |\psi(t)}}}.
\label{eq:QJ}
\end{equation}
Since the Bose-Hubbard Hamiltonian~\eqref{eq:FullHam} commutes with the total photon number operator $\hat N=\sum_{\ell}\hat n_\ell$, the no-jump evolution can be split into two independent parts, Hermitian evolution and damping at the rate $\gamma N$,
\begin{align}
    \ket{\psi_{\rm NJ}(t+\tau)} & =\frac{e^{-i \hat H^{(\gamma)}_{\rm NJ} \tau/\hbar }\ket{\psi(t)}}{\sqrt{\braket{\psi(t)|\big(e^{-i \hat H^{(\gamma)}_{\rm NJ} \tau/\hbar }\big)^\dagger e^{-i \hat H^{(\gamma)}_{\rm NJ} \tau/\hbar }|\psi(t)}}} \notag \\ & =e^{-\gamma N \tau}e^{-i \hat H_{\rm BH} \tau /\hbar }\ket{\psi(t)}, \label{eq:diss_NJ:evol}
\end{align}
assuming that one starts from a quantum state in a single photon-number sector with $N=\braket{\hat N}$. Physically this means that between the quantum jumps, the evolution of the system is identical to a one experiencing no dissipation. This conclusion holds even if we include the dephasing process since it induces no transitions between the photon-number sectors. Thus, if an experimental setting allows, by post-selecting based on the total number of bosons one can recover the quantum dynamics without any dissipation effects.

\begin{figure}[t]
\includegraphics[width=\columnwidth]{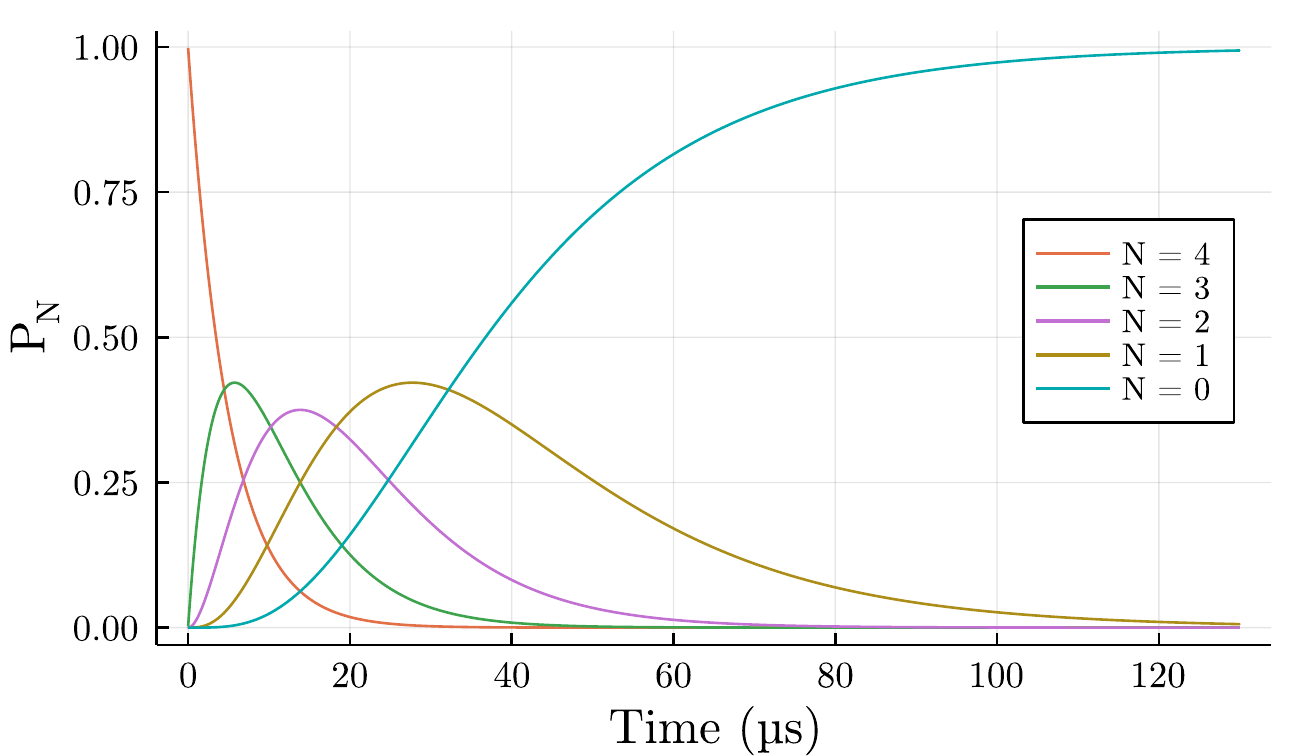}
\caption{The populations $P_N$ of the different photon number sectors as a function of time for $N=0,1,\ldots,4$, as given by Eq.~\eqref{eq:LDiss}. The system is initially in the photon number sector $N=4$ and the dissipation rate is $\gamma / 2 \pi \approx \SI{8}{\kilo\hertz}$ ($T_1 = \SI{20.0}{\micro\second}$).}
\label{fig:analytics}
\end{figure}

In order to solve the master equation~\eqref{eq:MEDiss}, we split the full density matrix into different photon number sectors, $\hat \rho=\sum_{N=0}^{N_{\rm max}} P_N(t)\hat\rho_N$, where $\hat \rho_N=\hat \Pi_N \hat \rho \hat \Pi_N$ and $\hat \Pi_N$ is a projector to the space of states with the total photon number $N$. The master equation then reduces to the rate equations $\dot P_N(t)=-\gamma N P_N(t)+\gamma (N+1)P_{N+1}(t)$ describing transitions between the blocks $N \to N-1$ at the rate $\gamma N$. By solving the rate equations, we obtain the probabilities
\begin{align}
P_N(t) =\binom{N_{\rm max}}{N}e^{-\gamma N t}(1-e^{-\gamma t})^{N_{\rm max}-N}
\label{eq:LDiss}
\end{align}
of being in the photon number sector $N$. These are depicted in Fig.~\ref{fig:analytics}.

The anharmonicity manifolds are spanned by the many-body Fock states $\ket{n_1, n_2, \ldots, n_L}=\ket{\bm{n}}$ with a fixed value for the anharmonicity $a=-\sum_\ell n_\ell(n_\ell-1)/2$ and the total photon number $N=\sum_\ell n_\ell$. For example, the lowest anharmonicity manifold is spanned by the states $\ket{N_\ell}$ where all bosons reside at a single site $\ell=1,2,\ldots,L$. The second-lowest anharmonicity manifold is spanned by the states $\ket{(N-1)_\ell, 1_k}$. Notice that in our shorthand notation of the many-body Fock states, we write explicitly only the boson numbers of the occupied sites. A photon-loss event by a quantum jump can lead to transitions between the anharmonicity manifolds. Note that the states in the lowest anharmonicity manifold are always mapped to a state in the lowest anharmonicity manifold of the next photon number manifold, $\ket{N_\ell}\to \ket{(N-1)_\ell}$. Similarly, the states in the highest anharmonicity manifold consisting of the hard-core boson states are mapped to the states in the highest anharmonicity manifold in the next photon number sector. In between, however, there are different decay channels depending on which site decays, although the total decay rate always equals to~$\gamma N$. As an example, the states $\ket{{N-1}_\ell, 1_k}$ decay to the manifold $\ket{{N-2}_\ell, 1_k}$ at a rate $\gamma (N-1)$ and to the manifold $\ket{(N-1)_\ell}$ at a rate~$\gamma$, see Fig.~\ref{fig:dndbhm-border-gradients-12-01-2022}(b).

The photon loss event, occurring at a single site, has generally a localizing effect. For example, if the state of the system is a superposition in the lowest anharmonicity manifold, $\ket{\psi}=\sum_{\ell=1}^L c_l\ket{N_\ell}$ with some coefficients $c_l$, then a quantum jump of Eq.~\eqref{eq:QJ} at the site $\ell'$ reduces it into a localized state $\ket{\psi}=\ket{N_\ell'}$, wiping out all the information on the coefficients $c_\ell$.

In addition to transitions, dissipation also causes decoherence. By considering the evolution of the off-diagonal term of the density matrix $\hat\rho$ between any many-body Fock states $\ket{n_1, n_2, \ldots, n_L}=\ket{\bm{n}}$ and $\ket{m_1, m_2, \ldots, m_L}=\ket{\bm{m}}$,
\begin{equation}
    \frac{d \braket{\bm{n}|\hat \rho|\bm{m}}}{d t}=\frac{\braket{\bm{n}|[\hat H,\hat{\rho}]|\bm{m}}}{i\hbar}-\frac{\gamma}{2}\sum_{\ell=1}^L(n_\ell +m_\ell)\braket{\bm{n}|\hat\rho|\bm{m}},
\end{equation}
we see that the decoherence rate $K^\gamma_{\bm{n},\bm{m}}$ between the states is simply proportional to the total number of photons they contain,
\begin{equation}
K^\gamma_{\bm{n},\bm{m}}=\frac{\gamma}{2}\sum_{\ell=1}^L(n_\ell+m_\ell). \label{eq.diss.decoher}
\end{equation}
In our case the system is limited to a single total photon number sector, and so the decoherence rate due to dissipation becomes
\begin{equation}
K^\gamma_{\bm{n},\bm{m}}=\gamma N.
\end{equation}
To summarize, dissipation in a transmon array is rather a simple process, leading essentially to a cascade $N \to N~-~1\to\ldots \to 0$ between the photon number sectors and to a loss of coherence as described by Eq.~\eqref{eq.diss.decoher}.

\section{Dephasing} \label{sec:dephasing}
The dephasing process originates from temporal fluctuations of the transmon frequencies, and it can be modeled using the master equation
\begin{equation}
    \frac{d \hat{\rho}}{d t} = - \frac{i}{\hbar}[ \hat{H}_{\rm BH}, \hat{\rho}] + \sum_{\ell=1}^{L} \frac{\kappa}{2} \left( 2 \hat{n}_{\ell} \hat{\rho} \hat{n}_{\ell} - \hat{n}^2_{\ell} \hat{\rho} - \hat{\rho} \hat{n}^2_{\ell} \right),
    \label{eq:MEDeph}
\end{equation}
with the jump operators $\sqrt{\kappa}\hat n_\ell$ for each transmon. We have here ignored the dissipation altogether for simplicity. By writing down the non-Hermitian Hamiltonian corresponding to the no-jump evolution of the dephasing process,
\begin{equation}
    \hat H_{\rm NJ}^{(\kappa)}=\hat H_{\rm BH}-i\sum_{\ell=1}^L\frac{\hbar \kappa}{2}\hat n^2_\ell,
\end{equation}
we see that the anti-Hermitian part $-i\sum_{\ell=1}^L\hbar \kappa\hat n^2_\ell/2$ does not commute with the Hermitian part $\hat H_{\rm BH}$. This implies that the dephasing process, and even its no-jump evolution generated by $\hat H_{\rm NJ}^{(\kappa)}$, induces transitions between the eigenstates of the attractive Bose-Hubbard Hamiltonian. Thus, in general, it is a more complicated process than the dissipation considered above.

\subsection{Decoherence of the many-body Fock states}
Let us first focus on the pure decoherence rates, introduced through the off-diagonal elements of the density matrix between many-body Fock states. Multiplication of Eq.~\eqref{eq:MEDeph} by a pair of arbitrary Fock states $\bra{\bm n}$ and $\ket{\bm{m}}$, with $\bm{n}\neq\bm{m}$, results in
\begin{align}
    \frac{d \braket{\bm{n}|\hat \rho|\bm{m}}}{d t}
    =&\frac{\braket{\bm{n}|[\hat H,\hat{\rho}]|\bm{m}}}{i\hbar}-\frac{\kappa}{2}\sum_{\ell=1}^L(n_\ell-m_\ell)^2\braket{\bm{n}|\hat \rho|\bm{m}}.
\end{align}
From the last term we see that the decoherence rate is given by
\begin{equation}
    K_{\bm{n},\bm{m}}^\kappa=\frac{\kappa}{2}|\bm{n}-\bm{m}|^2, \label{eq.deph.decoher}
\end{equation}
where $|\bm{n}|^2$ is the squared Euclidean norm of the vector~$\bm{n}=(n_1, n_2, n_3, \ldots, n_L)$.

Let us elucidate the effect of the decoherence to the dynamics of the attractive Bose-Hubbard model through four examples. First, in the lowest anharmonicity manifold, the state of the system is always some superposition of the $N$-boson stacks, $\ket{\psi}=\sum_{\ell} c_\ell \ket{N_\ell}$. Now, the decoherence process between the Fock states $\ket{N_\ell}$ and $\ket{N_{\ell'}}$ occurs at the rate of $K^\kappa_{\bm n,\bm m}=\kappa N^2$, see Fig.~\ref{fig:dephasing}(a). This means that the steady state of the system is a mixed state of $\ket{N_\ell}$, and it is reached at the rate of $\kappa N^2$. We study this in more detail in Sec.\ \ref{sec:disorder}.

Next, in the second-lowest anharmonicity manifold, the dynamics occurs between the states $\ket{(N-1)_\ell, 1_k}$. There are now a few possibilities for the speed of the decoherence process between the states $\ket{(N-1)_\ell, 1_k}$ and $\ket{(N-1)_{\ell'}, 1_{k'}}$, depending on the relative positions between the $N - 1$-boson stacks and the single bosons. The slowest rate is obtained when the stacks are located at the same site ($\ell' = \ell$). In this case, the decoherence occurs at the rate of $K^\kappa_{\bm{n},\bm{m}} = \kappa$. At the other extreme, if neither the stacks nor the bosons line up ($\ell', k' \neq \ell, k$), the decoherence rate is given by $K^\kappa_{\bm{n},\bm{m}} = \kappa [(N - 1)^2 + 1]$. Regardless of the different decoherence rates, the final steady state is a mixed state of $\ket{(N-1)_\ell, 1_k}$, reached on the time scales of~$\kappa^{-1}$.

As a third example, in the hard-core boson manifold, the lowest decoherence rate is attained between pairs of states that differ only by a single excitation pair, such as $\ket{\ldots0101\ldots}$ and $\ket{\ldots1001\ldots}$. For these, the decoherence rate is just $K^\kappa_{\bm{n},\bm{m}}=\kappa$. Correspondingly, the maximum decoherence rate is between the state pairs where the local excitation structure differs the most. For instance, when $N\leq L/2$, the maximum decoherence rate $K^\kappa_{\bm{n},\bm{m}}=\kappa N/2$ is achieved, for example, between the states $\ket{1010\ldots}$ and $\ket{0101\ldots}$.

Finally, in the same way that the dephasing leads to decoherence between states inside an anharmonicity manifold, it degrades coherence between states in different manifolds, see Fig.~\ref{fig:dephasing}(a).

To summarize, on the time scales of the order of $\kappa^{-1}$ or shorter, the density matrix is reduced to a diagonal matrix representing a mixed state within the initial anharmonicity manifold.

\subsection{Transitions between the anharmonicity manifolds}
The states in an anharmonicity manifold are actually weakly coupled to the states in the neighboring manifold via the hopping interaction. As derived in Ref.~\cite{Olli2021} through first-order non-degenerate perturbation theory in $J/U$, the $N$-boson stack state belonging to the anharmonicity manifold $a=-N(N-1)/2$ and the state of $N-1$ stacked bosons plus a lone boson belonging to the anharmonicity manifold $b=-(N-1)(N-2)$ are more accurately given by
    \begin{align}
    &\ket{N_\ell}_a=\ket{N_\ell}-\frac{J\sqrt{N}}{U(N-1)}\sum_{\sigma=\pm1}\ket{(N-1)_\ell, 1_{\ell+ \sigma }}, \label{eq:stack1} \\ 
    &\ket{(N-1)_\ell, 1_{\ell+ \sigma }}_b=\ket{(N-1)_\ell, 1_{\ell+ \sigma }}+\frac{J\sqrt{N}}{U(N-1)}\ket{N_\ell}.
\end{align}
Under unitary dynamics, the effect of this non-degenerate coupling is typically weak, and leads to fast oscillations between the two manifolds~\cite{Olli2022}. However, when combined with the dephasing process, it results in actual transitions between the manifolds. Intuitively, we can understand this by considering the state $\ket{N_\ell}_a$ of Eq.~\eqref{eq:stack1} under the action of the quantum jump operator $\sqrt{\kappa}\hat n_\ell$,
\begin{align}
   \ket{\psi^{(\kappa)}_{\rm QJ}}&=\frac{\sqrt{\kappa}\hat n_\ell \ket{N_\ell}_a}{\sqrt{\prescript{}{a}{\braket{N_\ell|\kappa \hat n_\ell^2|N_\ell}_a}}} \\
    &= \ket{N_\ell}_a+\frac{J\sqrt{N}}{U(N-1)}\sum_{\sigma=\pm1}\ket{(N-1)_\ell, 1_{\ell+ \sigma }}_b,\notag 
\end{align}
This holds to first order in $J/U$. The additional contribution by the states $\ket{(N-1)_\ell, 1_{\ell+ \sigma }}_b$ exemplifies the transitions between the anharmonicity manifolds. Similar phenomenon occurs for quantum jumps by operators $\sqrt{\kappa}\hat n_{\ell+\sigma}$. 

\begin{figure}
\includegraphics[width=\linewidth]{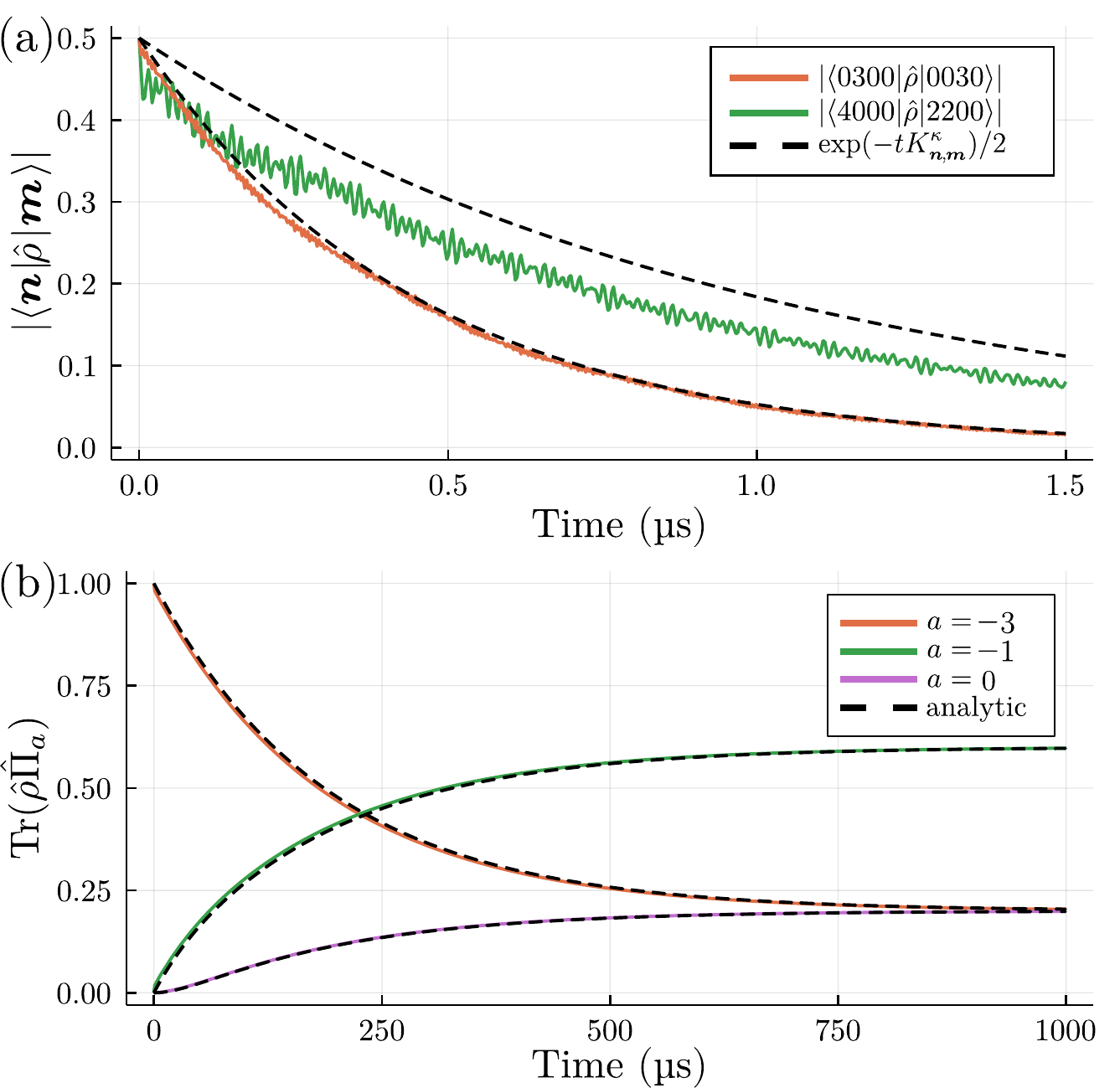}
\caption{(a) The decay of the off-diagonal elements $|\bra{0300} \hat{\rho} \ket{0030}|$ and $|\bra{4000} \hat{\rho} \ket{2200}|$ of the density matrix as a function of time $t$. The dashed lines show the comparison to the exponential decay $\exp (-t K^\kappa_{\bm n, \bm m}) / 2$, with the decoherence rates given by Eq.~\eqref{eq.deph.decoher}. (b) The populations $P_a=\Tr\left(\hat{\rho} \hat{\Pi}_a\right)$ of the anharmonicity manifolds as a function of time. The solution of the rate equation~\eqref{eq:ratePa} (dashed line) with the transition rates of Eq.~\eqref{eq:transition_rate} agrees very well with the numerical results. The parameters of the simulation are $J / 2\pi = \SI{20}{\mega\hertz}$, $U / 2\pi = \SI{230}{\mega\hertz}$, $\kappa / 2\pi \approx \SI{40}{\kilo\hertz}$ ($T^\star_2 = \SI{8}{\micro\second}$), $L=4$ and $N=3$. Here, we have ignored the dissipation altogether to highlight the dephasing phenomena.}
\label{fig:dephasing}
\end{figure}

For a more rigorous derivation, we first note that the dephasing reduces the density matrix to a diagonal form $\hat\rho=\sum_a P_a \pr_a$ on a fast time scale $\kappa^{-1}$. Here, $\pr_a$ is a projector to the anharmonicity manifold $a$ and the coefficient $P_a$ is the corresponding population. What we now consider are the slow transition rates between the anharmonicity manifolds. In other words, we consider the longer-time (slower) dynamics based on the concept of local equilibrium. We assume that the density matrix is always of the form determined by the equilibrium of the leading-order dynamics, but the relative weights between the different manifolds are different. Inserting this diagonal density matrix into the master equation~\eqref{eq:MEDeph} yields the rate equation
\begin{equation}
    \dot P_a=\sum_b(P_b-P_a)\Gamma_{ab}. \label{eq:ratePa}
\end{equation}
The effective transition rates $\Gamma_{ab}$ between the anharmonicity manifolds $a$ and $b$ are given by (see details on the derivation in App.~\ref{sec:app1})
\begin{align}
        \Gamma_{ab} & = \frac{\kappa}{\Tr(\pr_a) [\hbar U (a - b)]^2} \sum_{\ell=1}^L \Tr \left(\pr_a [\nop_\ell, \ham_J] \pr_b [\ham_J, \nop_\ell]\right) \notag \\
        & =  \frac{\kappa}{\Tr(\pr_a) [\hbar U (a - b)]^2} \sum_{\bm{n}_a, \bm{m}_b}|\bm{n}_a-\bm{m}_b|^2 |\braket{\bm{n}_a|\hat H_J|\bm{m}_b}|^2 
    \label{eq:transition_rate}
\end{align}
to first order in $J/U$, akin to the quantum jump consideration above. This implies that the dynamics between the anharmonicity manifolds occurs at rates of the order of $\kappa (J/U)^2$, assuming all the internal dynamics is fast compared to this.

As an example, let us again consider the lowest anharmonicity manifold $a_1 = -N(N - 1) /2$ spanned by the states $\ket{N_\ell}$. Applying $\hat{H}_J$ to any state in $a_1$ always gives us a state in the second-lowest manifold $b_1 = -(N - 1)(N - 2)/2$ spanned by the states $\ket{(N - 1)_\ell, 1_m}$. Thus, the only non-zero rate away from $a_1$ is
\begin{equation}
\Gamma_{a_1b_1} = 4 \kappa \left(\frac{J}{U} \right)^2 \frac{L - 1}{L} \frac{N}{(N - 1)^2}.
\label{eq:rate}
\end{equation}
Other transition rates can be derived similarly, see App.~\ref{sec:app1}. Figure~\ref{fig:dephasing}(b) compares the solution of the rate equation~\eqref{eq:ratePa} with the rates of Eq.~\eqref{eq:transition_rate} to the full numerical solution of the master equation, and demonstrates a very good agreement between the two approaches. Due to the $\kappa (J/U)^2$-dependence, the cooling and heating rates of Eq.~\eqref{eq:transition_rate} are slow with respect to the many-body decoherence and dissipation when calculated using realistic transmon array parameters. This is evident also by comparing the time scales of Fig.~\ref{fig:dephasing}(b) to those of Fig.~\ref{fig:dephasing}(a) or Fig.~\ref{fig:analytics}.

\subsection{Transitions out of the hard-core boson manifold}
Above, we focused on the lowest anharmonicity manifolds corresponding to the higher excited states of transmons. The formalism derived in Eqs.~\eqref{eq:ratePa}~--~\eqref{eq:transition_rate} also applies to the hard-core boson manifold, that is, the highest anharmonicity manifold with $a=0$. The hopping term of the Bose-Hubbard Hamiltonian couples the hard-core boson states to the states residing in the second-highest anharmonicity manifold $b=-1$. For example, the state $\ket{1101\ldots}$ is coupled to the state $\ket{0201\ldots}$. The dephasing-induced transition rate away from the hard-core boson manifold is
\begin{equation}
\Gamma_{\rm h.c.}=8\kappa \left(\frac{J}{U}\right)^2 \frac{N(N-1)}{L},\label{eq:hardcore}
\end{equation}
which interestingly increases quadratically as a function of the total excitation number $N$ in contrast to the transition rates in the lower end of the energy spectrum for the boson stacks. The transition rate of Eq.~\eqref{eq:hardcore} is the non-unitary counterpart of the always-on $ZZ$ interaction of hard-core bosons~\cite{Braumuller2022, Olli2022}. Both of them are corrections resulting from the exclusion of the higher excited states beyond the qubit subspace. Notice that the states with no neighboring excitations, such as $\ket{10101\ldots}$, experience no transitions, and the corresponding population does not leak out of the manifold.

\section{Effects of dissipation, dephasing, and disorder in boson stack dynamics}\label{sec:disorder}
In the absence of dissipation and dephasing, a characteristic property of the dynamics of a transmon array is the confinement of the state of the system into the anharmonicity manifold of the initial state~\cite{Olli2022}. As a consequence, an excited state of an individual transmon behaves like a quasiparticle which moves around the array without splitting into less-excited states. Its effective hopping rate is given by $\widetilde{J} = J[N/(N-1)!](J/U)^{N-1}$, where $N$ is the number of bosons comprising the quasiparticle, that is, the local occupation number. When multiple transmons in the array are highly excited, the dynamics can be further limited by effective interactions between the quasiparticles. Noting that dissipation and dephasing both cause transfer between the manifolds, it is worth looking into how they affect the quasiparticle dynamics. We will do this by studying two example cases in both of which the unitary dynamics is well described by the effective Hamiltonians of Ref.~\cite{Olli2022}.

\begin{figure}
    \centering
    \includegraphics[width=\linewidth]{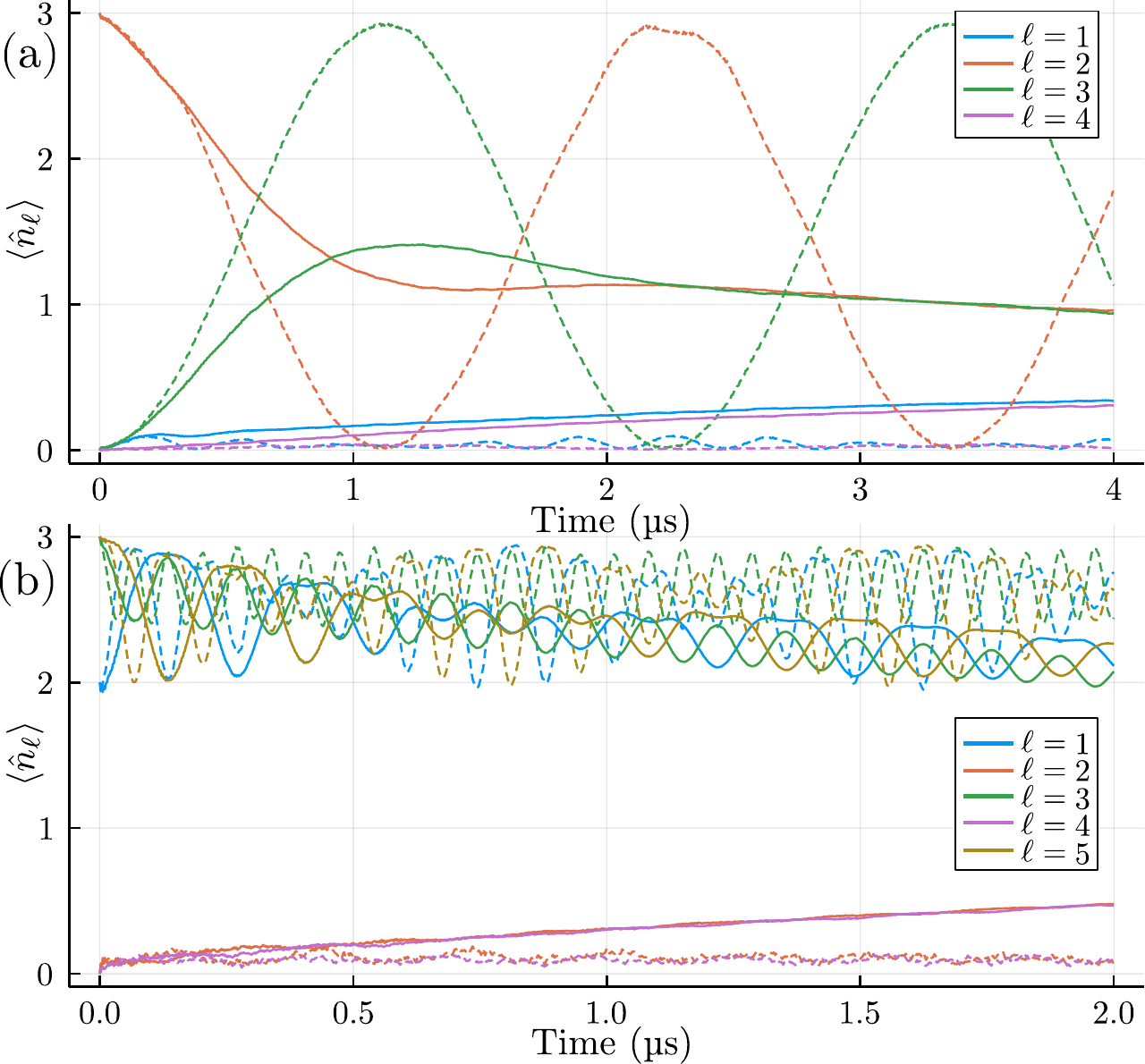}
    \caption{The local occupations $\braket{\nop_\ell}$ of the sites $\ell$ as a function of time. The non-unitary (solid) time-evolution as described by Eq.~\eqref{eq:MEF}, including both dissipation and dephasing, is compared against the unitary case (dashed) under the Bose-Hubbard Hamiltonian~\eqref{eq:FullHam}. The initial state is (a) a boson stack $\ket{3_2}$ with the total number of sites $L = 4$ and (b) an array of boson stacks $\ket{2_1, 3_3, 3_5}$ with $L = 5$. In both cases, the non-unitary dynamics is solved using the quantum trajectory approach, averaging over $16000$ trajectories. The parameters of the simulation are $J / 2\pi = \SI{20}{\mega\hertz}$, $U / 2\pi = \SI{230}{\mega\hertz}$, $\gamma / 2 \pi \approx \SI{8}{\kilo\hertz}$ ($T_1 = \SI{20.0}{\micro\second}$), and $\kappa / 2\pi \approx \SI{40}{\kilo\hertz}$ ($T^\star_2 = \SI{8}{\micro\second}$).}
    \label{fig:stacks}
\end{figure}

In our first example, we have a chain of $L = 4$ transmons with one of them prepared in the third exited level, so that $N = 3$. With the initial state $\ket{3_2}$, the unitary time-evolution results in oscillations along the array at the rate $\widetilde{J} = 3J (J/U)^2/2$. Due to the effective edge repulsion experienced by the quasiparticle, these oscillations are limited to only among the states $\ket{3_2}$ and $\ket{3_3}$. Now, with dephasing included, based on Eq.~\eqref{eq.deph.decoher} and the large distance between the states, we expect the oscillations to decay quite rapidly, at the rate $K^{\kappa}_{\bm n, \bm m}=\kappa N^2=9\kappa$ (on the time scales of $\approx\SI{0.4}{\micro\second}$). Figure~\ref{fig:stacks}(a) shows that this is indeed the case. We can also see that the small oscillations between the center sites and the edges allow dephasing to cause mixing within the entire initial anharmonicity manifold. Dissipation, which occurs on the time scales of $(3\gamma)^{-1} \approx \SI{7}{\micro\second}$, has only rather weak an effect on the time scales of the quasiparticle oscillations.

The second example we consider is a chain of $L = 5$ transmons with three quasiparticles, shown in Fig.~\ref{fig:stacks}(b). With the initial state $\ket{2_1, 3_3, 3_5}$, edge-localization and effective repulsive interactions between the quasiparticles lead to the unitary dynamics to be mostly limited to the subspace $\mathcal{S} = \{\ket{2_1, 3_3, 3_5}, \ket{3_1, 2_3, 3_5}, \ket{3_1, 3_3, 2_5} \}$. Here, the quasiparticle of two bosons moves along the chain via effective exchange interactions at the rate $\Xi=3 J (J/U) /4$. Within the subspace $\mathcal{S}$, the distances between the Fock vectors are much smaller than in the first example. Consequently, the decoherence rate due to dephasing inside $\mathcal{S}$ is considerably slower, $K^\kappa_{\bm n, \bm m}=\kappa$. The oscillations should therefore remain significantly longer. Note, however, that the distances between the subspace $\mathcal{S}$ and the rest of the initial anharmonicity manifold can be larger, and so mixing within the manifold might be faster.

Based on these examples, we can conclude that the effective many-body dynamics of the higher excited states of transmons can survive the presence of dissipation and dephasing at short time scales. However, even with a relatively large hopping frequency of $J/2\pi = \SI{20}{\mega\hertz}$, any dynamical effects occurring at third order or above in $J/U$ are too slow for the current analog simulators. Since the dissipation times are usually significantly larger than the dephasing times -- and one can, in principle, also remove the effects of dissipation by post-selecting based on the total boson number -- we conclude that the dephasing time sets the upper limit to the time scales available for operations involving the higher excited levels of transmons. Perhaps the easiest engineering approach to try and circumvent this problem is to increase the value of the hopping rate $J$ by enhancing the capacitive coupling between the transmons, either by using larger capacitors or by geometric means~\cite{Dalmonte15}. As the characteristic frequencies of the dynamics of the higher excited states scale as $(J/U)^N$, already a \SIrange{10}{25}{\percent} increase in the value of $J$ can have a drastic effect. 

\begin{figure}
\includegraphics[width=\linewidth]{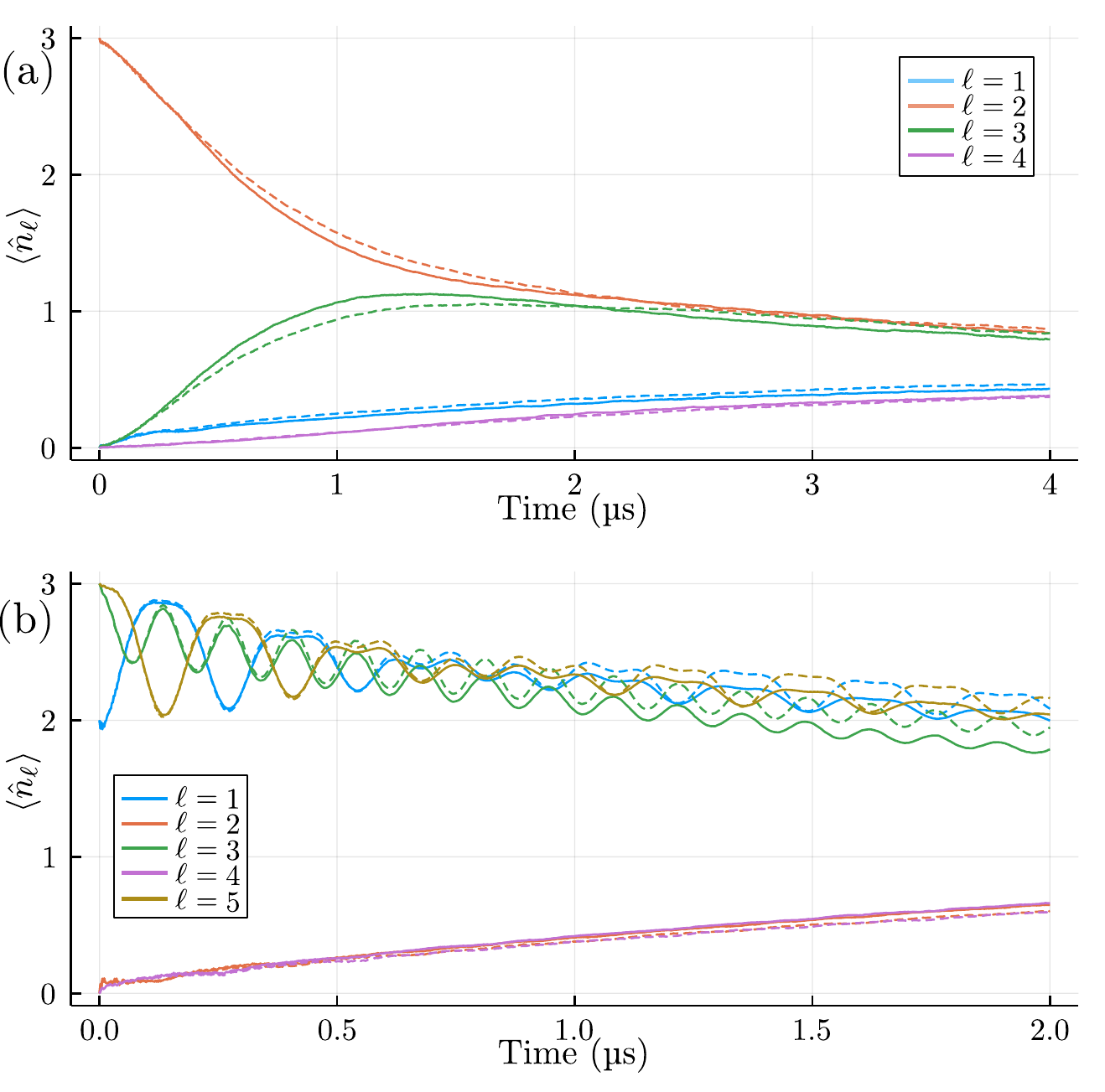}
\caption{The local occupations $\langle \nop_\ell \rangle$ of the sites $\ell$ as a function of time under non-unitary time-evolution with (solid) and without (dashed) disorder. In both cases, the mean values for the dissipation and dephasing rates are equal. Similar to Fig.~\ref{fig:stacks}, the initial state is (a) $\ket{3_2}$ (with $L = 4$) and (b) $\ket{2_1, 3_3, 3_5}$ (with $L = 5$). The simulation parameters are otherwise the same as in Fig.~\ref{fig:stacks}, except that the dissipation and dephasing times used in the disordered simulation are those of the qubits denoted (a) $Q10$--$Q13$ and (b) $Q10$--$Q14$ in Ref.~\cite{Zhu22}. This results in (a) $\gamma/2\pi=\SI{8.9}{\kilo\hertz}$, $\sigma_\gamma/2\pi=\SI{2.6}{\kilo\hertz}$, $\kappa/2\pi=\SI{111.5}{\kilo\hertz}$, $\sigma_\kappa/2\pi=\SI{48.9}{\kilo\hertz}$, and (b) $\gamma/2\pi=\SI{8.5}{\kilo\hertz}$, $\sigma_\gamma/2\pi=\SI{2.4}{\kilo\hertz}$, $\kappa/2\pi=\SI{111.2}{\kilo\hertz}$, $\sigma_\kappa/2\pi=\SI{43.8}{\kilo\hertz}$.}
\label{fig:disordered}
\end{figure}

\subsection{Parametric disorder in dissipation and dephasing}
Dissipation and dephasing in transmons arise from various sources, such as microscopic defects and fluctuators, quasiparticles, delicate device properties, and electromagnetic environments~\cite{Krantz19}. The degree to which these sources affect the dissipation and dephasing properties is mostly set by the device fabrication process, and thus cannot be much altered after the transmon array has been assembled. Since it is impossible to fabricate perfectly identical transmons, the transmon arrays exhibit rather strong parametric site-to-site disorder in the values of $\gamma_\ell$ and $\kappa_\ell$, the standard deviations being in the range of \SIrange[]{20}{50}{\percent}, see, e.g., Refs.~\cite{Ma2019, Arute2020, Gong21, Zhao22, Zhu22, Saxberg22}.

Let us briefly analyze the effects of dissipation disorder and dephasing disorder on the system by comparing to the ideal uniform case. Now, when the dissipation rates $\gamma_\ell$ are site-dependent, the anti-Hermitian part $-i\sum_{\ell}\hbar \gamma_\ell \hat a^\dagger_\ell \hat a_\ell$ in Eq.~\eqref{eq:diss_NJ} no more commutes with the Hermitian part~$\hat H_{\rm BH}$, and so the no-jump evolution is not just the Hermitian dynamics modified by the damping factor as was the case in Eq.~\eqref{eq:diss_NJ:evol}. Furthermore, the total rate of photon loss due to quantum jumps becomes time-dependent, $\braket{\psi(t)|\sum_\ell \gamma_\ell \hat a^\dagger_\ell \hat a_\ell^{}|\psi(t)}=\sum_\ell \gamma_\ell \braket{\hat n_\ell}(t)$. However, when the effective dynamics in the array is faster than $\gamma^{-1}$, with $\gamma$ being the site-averaged mean dissipation rate, then, by time-averaging, we see that the photon loss rate is still given by $\gamma N$. The decoherence rate between the many-body Fock states $\ket{\bm n}$ and $\ket{\bm m}$ due to the disordered dissipation becomes
\begin{equation}
    K^\gamma_{\bm n, \bm m}=\frac{1}{2}\sum_{\ell=1}^L\gamma_\ell (n_\ell+m_\ell)=\frac{1}{2}\bm{\gamma}\cdot(\bm{n}+\bm{m}),
\end{equation}
where we have defined the vector $\bm{\gamma}=(\gamma_1, \gamma_2, \ldots, \gamma_L)$. 

Similarly, the decoherence rate due to disordered dephasing becomes
\begin{equation}
    K^\kappa_{\bm n, \bm m}=\frac{1}{2}\sum_{\ell=1}^L\kappa_\ell (n_\ell-m_\ell)^2=\frac{1}{2} (\bm{n}-\bm{m}) \cdot \bm{\underline{\kappa}} \cdot (\bm{n}-\bm{m}),
\end{equation}
where we have defined the diagonal matrix $\bm{\underline{\kappa}} = \mathrm{diag}(\kappa_1, \kappa_2, \ldots, \kappa_L)$. The expression \eqref{eq:transition_rate} for the transition rates between the anharmonicity manifolds also generalizes to the case of disordered dephasing, see App.~\ref{sec:app1}. But since it considers transitions between uniformly distributed initial and final states in the manifolds, the averaged transition rate of Eq.~\eqref{eq:transition_rate} should describe also the disordered situation very well, with $\kappa$ now being the site-averaged mean dephasing rate.

In Fig.~\ref{fig:disordered}, we have numerically compared the non-unitary time-evolution with and without parametric disorder in dissipation and dephasing. To be experimentally as relevant as possible, the disorder pattern is taken from Ref.~\cite{Zhu22}. We see that, indeed, disorder produces only minor deviations in the case many sites are traversed within a coherence/dissipation time. If one studies frozen dynamics due to, for example, edge-localization, then disorder can naturally have notable effects. 

Finally we point out that the dephasing model of the master equation~\eqref{eq:MEF} is a simplification for the higher excited states in transmons. Equation~\eqref{eq.deph.decoher} yields that the dephasing rate between two consecutive higher exited states $\ket{n}$ and $\ket{n+1}$ in a single transmon would be independent on $n$: $K^\kappa_{n, n+1}=\kappa/2$. Due to increased sensitivity to charge fluctuations, the higher excited states of a transmon have worse than that coherence properties~\cite{Blok21, Peterer15}. Formally speaking, this phenomena can be accounted easily by replacing the operators $\sqrt{\kappa_\ell}\hat n_\ell$ with some other diagonal operators $\hat d_\ell$ in the dephasing part of the master equation~\eqref{eq:MEF}. For example, a possibility to model the enhanced decoherence can be $\hat d_\ell=\sqrt{\kappa_\ell}\exp[a_\ell (\hat n_\ell -1)]$  with a suitably chosen coefficients $\kappa_\ell$ and $a_\ell$ and by defining that $\hat d_\ell\ket{0}=0$. The results presented in Sec.~\ref{sec:dephasing} can be straightforwardly generalized for the operator $\hat d_\ell$. The detailed study of this topic is left a subject of future research.

\section{Conclusions} \label{sec:conc}
In this work, we studied non-unitary many-body dynamics in transmon arrays, taking into account dissipation and dephasing. Our focus was specifically on the dynamics of the higher excited states lying beyond the hard-core boson approximation, that is, we treated transmons as proper bosonic quantum multilevel systems in the experimentally relevant parameter regime of state-of-the-art devices. Instead of a local view to a single transmon, we investigated the non-unitary effects of dissipation and dephasing on global many-body states. In particular, we considered the many-body Fock states $\ket{\bm n}=\ket{n_1, n_2, \ldots, n_L}$ which can be grouped into different anharmonicity manifolds based on the values of the anharmonicity $A=-\sum_{\ell}n_\ell(n_\ell-1)/2$ and the total photon number $N=\sum_\ell n_\ell$.

The main findings demonstrated three clearly distinguishable processes: many-body decoherence, many-body dissipation, and transitions between the anharmonicity manifolds. The total decoherence rate between the many-body Fock states $\ket{\bm n}$ and $\ket{\bm m}$ is $K_{\bm n, \bm m} = \gamma N + \kappa |\bm n-\bm m|^2 / 2$. In the worst case, the decoherence rate scales as $N^2$. Furthermore, the dephasing rates $\kappa$ are typically an order of magnitude larger than the dissipation rates $\gamma$. Our numerical simulations with the dissipation rate $\gamma/2\pi\approx~\SI{8}{\kilo\hertz}~(T_1=\SI{20}{\micro\second}$) and the dephasing rate $\kappa/2\pi\approx \SI{40}{\kilo\hertz}~(T^\star_2=\SI{8}{\micro\second})$ show that the dynamics involving the higher excited states occurring on the time scales of $(J/U)^{-2} J^{-1}$ or less should be readily realizable with hopping rates $J/2\pi \lesssim \SI{20}{\mega\hertz}$. For higher-order many-body dynamics on the time scales of $(J/U)^{-3} J^{-1}$, the hopping rate $J$ needs to be modestly increased to the values of $\gtrsim 2\pi \times \SI{25}{\mega\hertz}$. The other two processes -- transitions between the photon number manifolds due to dissipation and heating/cooling transitions between the anharmonicity manifolds due to dephasing -- are both slower compared to the decoherence. To summarize, the dephasing times of the higher excited states put a practical limit on observing coherent many-body higher-excited-state dynamics in transmon arrays. 

As an outlook for the future, an interesting application to extend the present work is to study potential realizations of dynamical quantum phase transitions under non-unitary conditions. Moreover, since the dissipation and dephasing processes are inherently local, our results are readily applicable to general array geometries.

\section*{Acknowledgments} 
We thank Tuure Orell for useful discussions. The authors acknowledge financial support from the Academy of Finland (Grants No. 316619, No. 320086, No. 346035), the Kvantum Institute at the University of Oulu, and the Scientific Advisory Board for Defence (MATINE) of the Ministry of Defence of Finland.

\appendix
\section{Dephasing-induced transition rates between anharmonicity manifolds}\label{sec:app1}

We derive here in detail the expression for the transition rates between the anharmonicity manifolds by the combination of dephasing and unitary Bose-Hubbard dynamics. We work now in the Heisenberg picture with respect to the Hamiltonian $\hat H_{\rm BH}$ of Eq.~\eqref{eq:FullHam}, and denote operators in this picture as $\check{n}_\ell$ to distinguish them from the Schrödinger picture operators $\hat n_\ell$. The starting point is that the dephasing has rendered the density matrix fully diagonal, $\check{\rho} = \sum_a P_a \pr_a$. Here, the operators $\pr_a$ are projectors to the states in the anharmonicity manifold $a$ and $P_a$ is a coefficient describing the population in that manifold. Then, the master equation~\eqref{eq:MEDeph} can be written in the form of a rate equation
\begin{equation}
     \dot P_a = \sum_{b} (P_b - P_a) \Gamma_{ab},
\end{equation}
where the effective transition rate $\Gamma_{ab}$ from the anharmonicity manifold $a$ to the manifold $b$ is given by
\begin{equation}\label{eq:transition_rate:app}
    \Gamma_{ab} = \frac{1}{\Tr \pr_a} \sum_{\ell = 1}^L \kappa_\ell  \Tr\left [ \left(\pr_a \check{n}_\ell \pr_b\right)\left(\pr_b \check{n}_\ell \pr_a\right) \right],
\end{equation}
expressed in terms of the projectors $\pr_a$ and $\pr_b$ and allowing for disorder in the dephasing rates $\kappa_\ell$ for generality.

We can expand the Heisenberg-picture number operators in each anharmonicity manifold $a$ by considering the hopping Hamiltonian $\ham_J$ as a small perturbation,
\begin{equation}
\check{n}_\ell = \check{n}_\ell^{(0)} + \check{n}_\ell^{(1)} + \cdots,  
\end{equation}
with the zeroth-order and first-order terms in $J/U$ given by
\begin{align}
    \check{n}_\ell^{(0)} &= \sum_{aj,bk} e^{i(E_{aj} - E_{bk}) t / \hbar} \braE{a}{j}{0}  \nop_\ell \ketE{b}{k}{0} \ketE{a}{j}{0} \braE{b}{k}{0},  \\
    \check{n}_\ell^{(1)} &= \sum_{aj, bk} e^{i(E_{aj} - E_{bk}) t / \hbar}  \left ( \braE{a}{j}{1} \nop_\ell \ketE{b}{k}{0} \ketE{a}{j}{0} \braE{b}{k}{0} \right. \notag \\
    &\hspace{40pt}+ \braE{a}{j}{0} \nop_\ell \ketE{b}{k}{1} \ketE{a}{j}{0} \braE{b}{k}{0} \notag \\
    &\hspace{40pt}+ \braE{a}{j}{0} \nop_\ell \ketE{b}{k}{0} \ketE{a}{j}{1} \braE{b}{k}{0} \notag \\
    &\hspace{40pt}+ \left. \braE{a}{j}{0} \nop_\ell \ketE{b}{k}{0} \ketE{a}{j}{0} \braE{b}{k}{1} \right) .
\end{align}
Here, $E_{a j}$ and $\ket{E_{a j}}$ are the eigenenergies and the corresponding eigenstates of the Bose-Hubbard Hamiltonian~\eqref{eq:FullHam}, and we have further expanded the states in powers of $J/U$ as $\ket{E_{a j}} = \ketE{a}{j}{0} + \ketE{a}{j}{1} + \ldots$, see also Ref.\ \cite{Olli2022}. Now, the state $\ketE{a}{j}{0}$ belongs to the anharmonicity manifold $a$. Since different manifolds are orthogonal to each other, and operating with $\nop_\ell$ keeps us within a manifold, we always have $\braE{a}{j}{0} \nop_\ell \ketE{b}{k}{0} = 0$ for $b \neq a$. Thus, multiplying the above equations with projectors from both sides, we obtain
\begin{align}
    \pr_a \check{n}_\ell^{(0)} \pr_b &= 0, \\
    \pr_a \check{n}_\ell^{(1)} \pr_b &= \sum_{j, k} e^{i(E_{aj} - E_{bk}) t / \hbar} \ketE{a}{j}{0} \braE{b}{k}{0}  \\
    &\hspace{30pt}\times \left(\braE{a}{j}{1} \nop_\ell \ketE{b}{k}{0} + \braE{a}{j}{0} \nop_\ell \ketE{b}{k}{1} \right). \notag
\end{align}
This means that the transition rates $\Gamma_{a b}$ of Eq.~\eqref{eq:transition_rate:app} are second order in $J/U$, for
\begin{align}
    &\left(\pr_a \check{n}_\ell \pr_b \right) \left(\pr_b \check{n}_\ell \pr_a \right) \approx \left(\pr_a \check{n}_\ell^{(1)} \pr_b \right) \left(\pr_b \check{n}_\ell^{(1)} \pr_a \right)  \\ 
    &\qquad= \sum_{jkm} e^{i(E_{aj} - E_{am}) t / \hbar} \ketE{a}{j}{0} \braE{a}{m}{0} \notag \\
    &\hspace{55pt}\times \left(\braE{a}{j}{1} \nop_\ell \ketE{b}{k}{0} + \braE{a}{j}{0} \nop_\ell \ketE{b}{k}{1} \right) \notag \\
    &\hspace{55pt}\times \left(\braE{b}{k}{1} \nop_\ell \ketE{a}{m}{0} + \braE{b}{k}{0} \nop_\ell \ketE{a}{m}{1} \right) \notag,
\end{align}
where we used the orthonormality of the zeroth-order eigenstates to eliminate one of the sums. Taking the trace removes the time-dependent exponential factor, yielding
\begin{align}
   &\Tr\left[ \left(\pr_a \check{n}_\ell \pr_b \right) \left(\pr_b \check{n}_\ell \pr_a \right) \right] \\
   &\qquad\approx \sum_{jk} \left(\braE{a}{j}{1} \nop_\ell \ketE{b}{k}{0} + \braE{a}{j}{0} \nop_\ell \ketE{b}{k}{1} \right) \notag \\
   &\hspace{55pt} \times \left(\braE{b}{k}{1} \nop_\ell \ketE{a}{j}{0} + \braE{b}{k}{0} \nop_\ell \ketE{a}{j}{1} \right) \notag.
\end{align}
Using again the orthogonality of the different anharmonicity manifolds, we do not need to know the components of $\ketE{a}{j}{1}$ lying in $a$, but only their projections
\begin{equation}
    \pr_b \ketE{a}{j}{1} = \frac{\pr_b \ham_J}{\hbar U (a - b)} \ketE{a}{j}{0}
\end{equation}
to the manifold $b$. Substituting these into the equation above, and noting that $\sum_k \ketE{b}{k}{0}\braE{b}{k}{0} = \pr_b$ and $\sum_j \braket{E_{a j}^{(0)} | \hat{O} | E_{a j}^{(0)}} = \Tr[\pr_a \hat{O} \pr_a]$, we can write the leading-order approximation for the effective transition rates as
\begin{align}
    \Gamma_{ab} &= \sum_{\ell=1}^L \kappa_\ell \frac{\Tr \left(\pr_a [\ham_J, \nop_\ell] \pr_b [\ham_J, \nop_\ell] \pr_a\right)}{(\Tr\pr_a) [\hbar U (a - b)]^2} \label{eq:app:transrate}\\
    &=  \frac{1}{\Tr\pr_a} \sum_{\bm{n}_a, \bm{m}_b}\left[\frac{\bm{\kappa}\cdot(\bm{n}_a-\bm{m}_b)}{\hbar U (a-b)}\right]^2 |\braket{\bm{n}_a|\hat H_J|\bm{m}_b}|^2. \notag
\end{align}

In the main text, we give the transition rate from the lowest to the second-lowest anharmonicity manifold in the case of uniform dephasing, see Eq.~\eqref{eq:rate}. Equation~\eqref{eq:app:transrate} is simple enough to allow us to compute explicitly transition rates also between other anharmonicity manifold pairs. For example, from the second-lowest anharmonicity manifold spanned by the states $\ket{(N - 1)_\ell, 1_m}$, we can get to three different manifolds using the hopping Hamiltonian $\hat H_{\rm J}$: (i) to the lowest anharmonicity manifold $a_1 = -N(N - 1)/2$ spanned by the states $\ket{N_\ell}$; (ii) to the manifold $b_2 = -(N - 2)(N - 3)/2 + 1$, spanned by the states $\ket{(N - 2)_\ell, 2_m}$; and (iii) to the manifold $b_3 = -(N - 2)(N - 3)/2$ spanned by the states $\ket{(N - 2)_\ell, 1_m, 1_n}$. The corresponding rates are 
\begin{align}
\Gamma_{b_1a_1} &= 4 \kappa \left(\frac{J}{U} \right)^2 \frac{N}{L(N - 1)^2}, \label{eq:rateOne}\\
\Gamma_{b_1b_2} &= 8 \kappa \left(\frac{J}{U} \right)^2 \frac{N - 1}{L(N - 3)^2}, \label{eq:rateTwo}\\
\Gamma_{b_1b_3} &= 4 \kappa \left(\frac{J}{U} \right)^2 \frac{L - 2}{L} \frac{N - 1}{(N - 2)^2}.\label{eq:rateThree}
\end{align}
The expression~\eqref{eq:app:transrate} is also relatively simple to be used numerically to compute transition rates between arbitrary anharmonicity manifold pairs.

\newpage

\bibliography{references}

\end{document}